\documentclass[12pt,a4paper]{article}
\usepackage{a4wide}
\usepackage{amsmath}
\usepackage{amssymb}
\usepackage{epsfig}
\usepackage{subfigure}
\usepackage{exscale}
\usepackage{float}
\usepackage{bbm}
\usepackage[numbers,sort&compress]{natbib}
\usepackage{pst-plot, pstricks,pst-math}
\usepackage{fancybox,amssymb,color}
\usepackage{graphicx}
\usepackage{pstricks, color, graphicx, epsfig, psfrag}
\usepackage{amsfonts,amsmath,amssymb,slashed}
\usepackage{dsfont}
\usepackage{bbm,bm}
\usepackage{fancyhdr, a4wide}
\usepackage[english]{babel}
\usepackage{subfigure}

\makeatletter
\@addtoreset{equation}{section}
\makeatother

\title{Low-Temperature Properties of Ferromagnetic Spin Chains in a Magnetic Field}

\author{Christoph P.\ Hofmann$^a$ \\ \\
\normalsize {$^a$ Facultad de Ciencias, Universidad de Colima} \\
\vspace{0.3cm}
\normalsize {Bernal D\'iaz del Castillo 340, Colima C.P.\ 28045, Mexico} \\}

\begin{document}

\maketitle

\begin{abstract} \normalsize

\end{abstract}

The thermodynamic properties of ferromagnetic spin chains have been analyzed with a variety of microscopic methods over the years: Bethe
ansatz, spin-wave theory, Schwinger-boson mean-field theory, Green functions and renormalization group methods. Surprisingly, in all these
different studies, to the best of our knowledge, the manifestation of the spin-wave interaction in the low-temperature series for the
thermodynamic quantities has been ignored. In the present work, we address this problem by following a different path, based on the
systematic effective Lagrangian method. We evaluate the partition function up to two-loop order and derive the low-temperature expansion
of the energy density, entropy density, heat capacity, magnetization and susceptibility in the presence of a weak external magnetic field.
Remarkably, the spin-wave interaction only manifests itself beyond two-loop order. In particular, there is no term of order $T^2$ in the
low-temperature series of the free energy density. This is the analog of Dyson's statement that, in the case of three-dimensional ideal
ferromagnets, there is no term of order $T^4$ in the low-temperature series of the free energy density. The range of validity of our
series is critically examined in view of the Mermin-Wagner theorem. We also compare our results with the condensed matter literature and
point out that there are some misleading statements.


\maketitle

\section{Introduction}
\label{Intro}

Ideal ferromagnets, i.e., ferromagnetic systems which are governed by purely isotropic exchange coupling between nearest neighbors and by
the interaction with a weak external magnetic field, have been the subject of an impressive number of publications over the past few
decades. In three spatial dimensions, the situation is well-known: after various unsuccessful attempts, the correct low-temperature series
for the spontaneous magnetization was first given by Dyson in Ref.~\citep{Dys56}. Many authors after Dyson also discussed the
low-temperature series for the three-dimensional ideal ferromagnet, based on other microscopic methods, such as spin-wave theory and
Green functions. A simple and elegant method, according to Dyson \citep{Dys96}, is provided by Ref.~\citep{Zit65}. More recently, within
the systematic effective Lagrangian method, Dyson's series was rederived in Ref.~\citep{Hof02} and extended to higher orders in
Ref.~\citep{Hof11a}.

Remarkably, regarding the low-temperature series describing two-dimensional ferromagnets, only a few papers are available, all of them
dealing with noninteracting spin waves \citep{ML69,Col72,YK73,Tak86,Tak87a,Tak87b,Tak90,AA90a,AA90b,Yab91,SSI94,NT94,PM98,KSK03}.
Within the effective Lagrangian framework, the question of how the spin-wave interaction manifests itself in the low-temperature
properties of two-dimensional ideal ferromagnets has been solved in Refs.~\citep{Hof12a,Hof12b}. 

In the present work, we apply the effective Lagrangian method to ferromagnetic spin chains -- it is the first time, to the best of our
knowledge, that one-dimensional systems are studied within the systematic effective loop expansion. As we will explain below, Lorentz- or
Pseudo-Lorentz-invariant systems, such as antiferromagnets, cannot be systematically analyzed within the framework of effective
Lagrangians in one spatial dimension: the linear, i.e., relativistic, dispersion relation of the magnons in an antiferromagnet spoils the
systematic loop expansion where the method is based upon. In this respect ferromagnetic magnons, which display a quadratic dispersion
relation, represent an interesting exception: for this nonrelativistic system the loop expansion perfectly works in one spatial dimension,
such that the powerful method of effective Lagrangians can indeed be applied to ferromagnetic spin chains.

The effective Lagrangian method corresponds to an expansions of observables in powers of momentum or, equivalently, in powers of
temperature. The systematic effective framework is based upon the fact that loops in Feynman diagrams are suppressed by some power $n$ of
momentum -- otherwise the loop expansion does not converge and the effective field theory method fails. As we will see, the power $n$
referring to the suppression of momentum, depends on the spatial dimension of the system and on the dispersion relation. In
(Pseudo-)Lorentz-invariant effective field theories, which include the effective theories of quantum chromodynamics and antiferromagnets,
the Goldstone bosons display a linear, i.e., relativistic, dispersion relation. Here, every loop in a Feynman diagram corresponds to a
suppression of $p^{d_s-1}$ powers of momentum. The effective expansion thus works in three and two spatial dimensions, but is not
applicable to one-dimensional (Pseudo-)Lorentz-invariant systems . In this respect the ferromagnet, its magnon displaying a quadratic
dispersion relation, represents a peculiar case. Here, every loop in a Feynman diagram leads to a suppression of $p^{d_s}$ powers of
momentum, implying that the systematic effective Lagrangian method works in three, two and one spatial dimension.

In the present study, we evaluate the partition function of ferromagnetic spin chains up to two-loop order in the presence of a weak
external magnetic field. The low-temperature series for the free energy density, energy density, entropy density, heat capacity,
magnetization and susceptibility are given. It is pointed out that the spin-wave interaction does not yet manifest itself at this order of
the effective expansion -- it only enters at the three-loop level.

The range of validity of the low-temperature series is more restricted in one than in two spatial dimensions. This has to do with the fact
that, unlike in two spatial dimensions, the nonperturbatively generated correlation length of ferromagnetic magnons no longer is
exponentially large. We carefully examine the domain of validity of the effective low-temperature series and, in particular, underline
that it is conceptually inconsistent to switch off the magnetic field in these expressions.

While the thermodynamics of ferromagnetic spin chains has not been analyzed with effective Lagrangians so far, these systems have
attracted a lot of attention over the past few decades and many methods have been used to derive their low-temperature properties. Early
studies were based on the Bethe ansatz, amounting to numerically solving a system of coupled integral equations
\citep{Tak71,Tak73,Sch85,TY85,YT86,Sch86,LS87,Yam90,GBT07}.
Later on, modified spin-wave theory -- a variant of conventional spin-wave theory, designed to cope with two- and one-dimensional systems
-- was used in Refs.~\citep{Tak86,Tak87a,SB06}.
Ferromagnetic spin chains were also addressed with Schwinger-boson mean-field theory
\citep{AA90a,AA90b},
Green functions \citep{SSI94,KY72,CAEK89,HCHB01,JIRK04,APPC08,JIBJ08,LCSWD11},
spin-wave theory at constant order parameter \citep{KSK03},
renormalization group and scaling methods \citep{Kop89,CAEK89,NaT94,NHT95a,NHT95b,RS95,TNS96,Sac06},
and by Monte Carlo simulations \citep{CL83,Lyk83,FU86,CCL87,DL90,SSC97,GPL05,JIBJ08}.
Yet other approaches to ferromagnetic spin chains can be found in Refs.~\citep{ZE90,YG99,CTVV00,CCC01,DK06,DK12}.

Most of these studies focus on the limit of a zero magnetic field. Our effective analysis, on the other hand, is valid in a different
regime where the magnetic field is weak, but not zero. Still, some of the above authors also consider the case of a nonzero magnetic
field, such that their findings can be compared with our effective results. As it turns out, there are some misleading statements in the
literature regarding conventional and modified spin-wave theory.

We would like to stress again that the manifestation of the spin-wave interaction in the low-temperature series describing ferromagnetic
spin chains has not been considered explicitly in any of the above references. So it remains rather unclear whether the low-temperature
series presented in these studies are indeed correct, i.e., complete up to the order considered, or whether they receive corrections due
to the spin-wave interaction. This is one of the main problems we will address in the present work.

The rest of the paper is organized as follows. In Sec.~\ref{Loops} we provide a brief outline of the effective Lagrangian method with
special attention to the loop counting in one spatial dimension. The partition function for ferromagnetic spin chains is evaluated
perturbatively up to two-loop order and the low-temperature series for various thermodynamic quantities are derived in Sec.~\ref{Results}.
The range of validity of these series is critically examined in Sec.~\ref{MerminWagner}. The relevant condensed matter literature on
ferromagnetic spin chains is reviewed and compared with our effective results in Sec.~\ref{CDM}. Finally, Sec.~\ref{Summary} contains our
conclusions.

At the end of this section, we would like to mention that the systematic and model-independent effective Lagrangian method has been used
to study a variety of condensed matter systems with a spontaneously broken internal spin symmetry. In three spatial dimensions, the
low-energy properties of ferromagnets and antiferromagnets were analyzed in
Refs.~\citep{Leu94a,Hof99a,Hof99b,RS99a,RS99b,RS00,Hof01,Hof11c}. Two-dimensional ferromagnets and antiferromagnets were the subject of
Refs.~\citep{Hof12a,Hof12b,HL90,HN91,HN93,Hof10}. Of particular interest are two-dimensional antiferromagnets which turn into
high-temperature superconductors upon doping with either holes or electrons. These systems have been analyzed within the effective field
theory framework, both for underlying square and honeycomb lattices,
in Refs.~\citep{KMW05,BKMPW06,BKPW06,BHKPW07,BHKMPW07,BHKPW08,JKHW09,BHKMPW09,Hof11b,KBWHJW12,VHJW12}. Finally, the consistency of the
effective Lagrangian method with high-precision numerical simulations and microscopic models was demonstrated in
Refs.~\citep{WJ94,GHJNW09,JW11,GHJPSW11,GHKW10}.

\section{Effective Lagrangians and Loop Counting}
\label{Loops}

In this section, we will focus on some essential aspects of the effective Lagrangian method at finite temperature. The interested reader
may find a more detailed account on finite-temperature effective Lagrangians in appendix A of Ref.~\citep{Hof11a} and in the various
references given therein. In addition, for pedagogic introductions to the effective Lagrangian technique, we refer to
Refs.~\citep{Bur07,Brau10,Leu95,Sch03,Goi04}.

The basic degrees of freedom of the effective Lagrangian are the Goldstone bosons which are a consequence of the spontaneously broken
continuous symmetry. At low energies or low temperatures these particles dominate the physical behavior of the system. In the present case
of ferromagnetic spin chains, we are dealing with magnons which are the  Goldstone bosons of the spontaneously broken spin rotation
symmetry: while the Heisenberg model is invariant under the group O(3), the ground state -- at zero temperature -- is only invariant under
O(2).

The systematic construction of the terms in the effective Lagrangian is straightforward: the link between the underlying theory and the
effective theory is provided by the symmetries \citep{Wei79,GL85,Leu94b}. One first identifies all symmetries of the underlying theory. In
our case, the Heisenberg model exhibits an O(3) spin rotation symmetry, as well as parity and time reversal symmetry. The effective
Lagrangian, or more precisely, the effective action for the ferromagnetic spin chain,
\begin{equation}
{\cal S}_{eff} = \int \! d^2 x \, {\cal L}_{eff} \, ,
\end{equation}
inherits all these symmetries of the underlying Heisenberg model.

The various terms in the effective Lagrangian can be organized according to the number of space and time derivatives which act on the
Goldstone boson fields. At low energies or momenta, terms which contain only a few derivatives are the dominant ones, while terms with
more derivatives are suppressed. This organization of terms is the basis for the systematic expansion of quantities of physical interest
in powers of momentum $p$.

For the ideal ferromagnet in $d_s$ spatial dimensions, the leading-order effective Lagrangian is of order $p^2$ and takes the form
\citep{Leu94a}:
\begin{equation}
\label{leadingLagrangian}
{\cal L}^2_{eff} = \Sigma \frac{\epsilon_{ab} {\partial}_0 U^a U^b}{1+ U^3}
+ \Sigma \mu H U^3 - \mbox{$\frac{1}{2}$} F^2 {\partial}_r U^i {\partial}_r U^i \, , \qquad r = x_1, \, \dots , x_{d_s} \, .
\end{equation}
The effective degrees of freedom are the two real components of the magnon field, $U^a (a=1,2)$, which represent the first two components
of the three-dimensional magnetization unit vector $U^i = (U^a, U^3)$, transforming with the vector representation of the rotation group.
The quantity $H$ is the magnetic field which points along the third direction, ${\vec H} = (0,0,H)$. While the structure of the above
terms is unambiguously determined by the symmetries of the underlying theory, at this order, we have two a priori unknown low-energy
constants: the spontaneous magnetization at zero temperature $\Sigma$ and the constant $F$. These low-energy couplings have to be
determined by experiment, numerical simulation or by a comparison with the microscopic theory.

The above Lagrangian leads to a quadratic dispersion relation,
\begin{equation}
\omega({\vec k}) = \gamma {\vec k}^2 + {\cal O}( { |{\vec k}| }^4) \, , \quad \gamma \equiv \frac{F^2}{\Sigma} \, ,
\end{equation}
characteristic of ferromagnetic magnons. It is important to note that this relation dictates how we have to count time and space
derivatives in the systematic effective expansion: One time derivative (${\partial}_0$) is on the same footing as two space derivatives
(${\partial}_r {\partial}_r$), i.e., two powers of momentum count as only one power of energy or temperature: $k^2 \propto \omega, T$.

As derived in Ref.~\citep{Hof02}, the next-to-leading terms in the effective Lagrangian are of order $p^4$ and contain four spatial
derivatives. In two or three spatial dimensions we have a total of three independent terms,
\begin{equation}
{\cal L}^4_{eff} = l_1 {( {\partial}_r U^i {\partial}_r U^i )}^2 + l_2 {( {\partial}_r U^i {\partial}_s U^i )}^2 + l_3 \Delta U^i \Delta U^i
\qquad (d_s = 2,3) \, .
\end{equation}
Here $\Delta$ denotes the Laplace operator in $d_s$ spatial dimensions. The next-to-leading order effective Lagrangian involves the three
effective coupling constants $l_1, l_2$ and $l_3$. In one spatial dimension, however, the first two terms coincide and we are left with
only two independent terms of order $p^4$,
\begin{equation}
\label{Leff4}
{\cal L}^4_{eff} = l_1 {( {\partial}_{x_1} U^i \, {\partial}_{x_1} U^i )}^2
+ l_3 \, {\partial}_{x_1}^2 U^i \, {\partial}_{x_1}^2  U^i \qquad (d_s = 1) \, .
\end{equation}
Higher-order pieces ${\cal L}^6_{eff}, {\cal L}^8_{eff}, \dots$ of the effective Lagrangian, as we will discuss below, are irrelevant for
the evaluation of the partition function considered in this work.

In finite-temperature field theory the partition function is represented as a Euclidean functional integral
\begin{equation}
\label{TempExp}
\mbox{Tr} \, [\exp(- {\cal H}/T)] = \int [dU] \, \exp \Big(- {\int}_{\! \! \! {\cal T}} \! \! d^{d_s+1}x \, {\cal L}_{eff}\Big) \, .
\end{equation}
The integration extends over all magnon field configurations which are periodic in the Euclidean time direction
$U({\vec x}, x_4 + \beta) = U({\vec x}, x_4)$, with $\beta \equiv 1/T$. The quantity ${\cal L}_{eff}$ on the right-hand side is the
Euclidean form of the effective Lagrangian, which consists of a string of terms
\begin{equation}
{\cal L}_{eff} = {\cal L}^2_{eff} + {\cal L}^4_{eff} + {\cal O}(p^6) \, ,
\end{equation}
involving an increasing number of space and time derivatives.

\begin{figure}
\includegraphics[width=14.7cm]{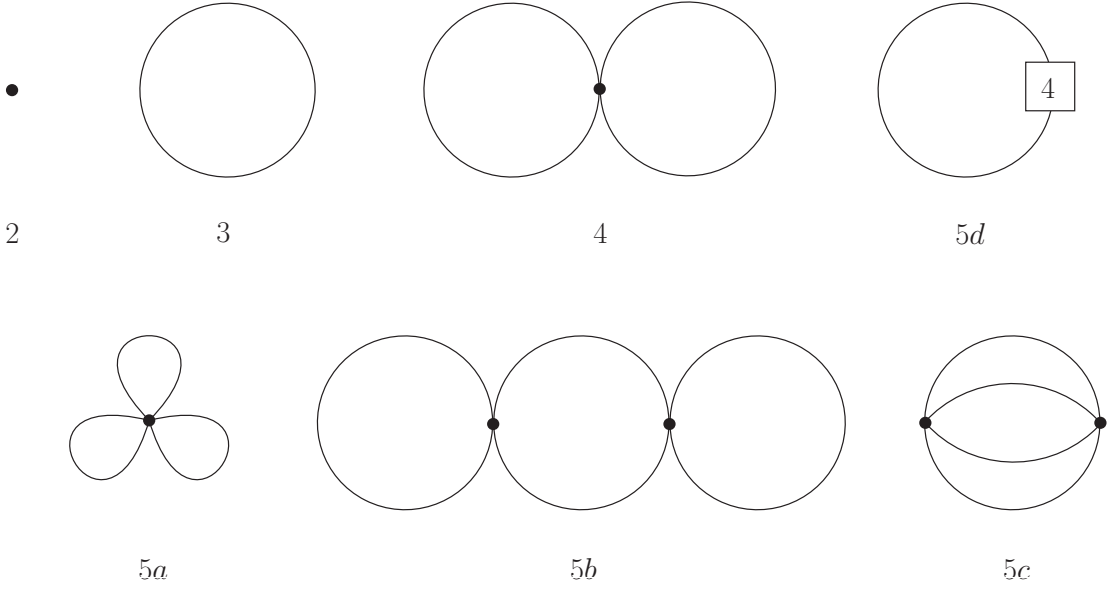}

\caption{Feynman graphs related to the low-temperature expansion of the partition function for the ferromagnetic spin chain up to order
$p^5$. The numbers attached to the vertices refer to the piece of the effective Lagrangian they come from. Vertices associated with the
leading term ${\cal L}^2_{eff}$ are denoted by a filled circle. Note that ferromagnetic loops are suppressed by one power of momentum in
one spatial dimension, $d_s$=1.}
\label{figure1}
\end{figure}

The virtue of the representation (\ref{TempExp}) lies in the fact that it can be evaluated perturbatively. To a given order in the
low-temperature expansion only a finite number of Feynman graphs and only a finite number of effective coupling constants contribute. The
low-temperature expansion of the partition function is obtained by considering the fluctuations of the spontaneous magnetization vector
field ${\vec U} = (U^1, U^2, U^3 )$ around the ground state ${\vec U_0} = (0, 0, 1)$, i.e., by expanding $U^3$ in powers of the spin-wave
fluctuations $U^a$,
\begin{equation}
U^3 = \sqrt{1-U^aU^a} = 1 - \mbox{$\frac{1}{2}$} U^a U^a - \mbox{$\frac{1}{8}$} U^a U^a U^b U^b - \dots \, . 
\end{equation}
Inserting this expansion into formula (\ref{TempExp}) one then generates the set of Feynman diagrams illustrated in Fig.~\ref{figure1}.
The leading contribution in the exponential on the right-hand side of Eq.~(\ref{TempExp}) is of order $p^2$ and originates from
${\cal L}^2_{eff}$. It contains a term quadratic in the spin-wave field $U^a$ -- with the appropriate derivatives and the magnetic field
displayed in Eq.(\ref{leadingLagrangian}) -- and describes free magnons. The corresponding diagram for the partition function is the
one-loop diagram 3 of Fig.~\ref{figure1}.

The remainder of the effective Lagrangian in the path integral formula for the partition function (\ref{TempExp}), i.e.,
${\cal L}^4_{eff} + {\cal L}^6_{eff} + \dots$, is treated as a perturbation. The Gaussian integrals are evaluated in the standard manner
(see Ref.~\citep{Kap89}, in particular chapter 3), and one arrives at a set of Feynman rules which differ from the zero-temperature rules
of the effective Lagrangian method only in one respect: the periodicity condition imposed on the magnon fields modifies the propagator. At
finite temperature, the propagator is given by
\begin{equation}
\label{ThermalPropagator}
G(x) = \sum_{n \,= \, - \infty}^{\infty} \Delta({\vec x}, x_4 + n \beta) \, , \qquad {\vec x} = (x_1, \dots, x_{d_s}) \, ,
\end{equation}
where $\Delta(x)$ is the Euclidean propagator at zero temperature,
\begin{eqnarray}
\label{Propagator}
\Delta (x) & = & \int \! \frac{d k_4 \, d^{d_s}\!k}{(2\pi)^{d_s+1}}
\frac{e^{i{\vec k}{\vec x} - i k_4 x_4}}{\gamma {\vec k}^2 - i k_4 + \mu H} \nonumber \\
& = & \Theta(x_4) \int \! \frac{d^{d_s}\!k}{(2\pi)^{d_s}} \, e^{i{\vec k}{\vec x} - \gamma {\vec k}^2 x_4 -\mu H x_4} \, . 
\end{eqnarray}
An explicit representation for the thermal propagator, dimensionally regularized in the spatial dimension
$d_s$, is
\begin{equation}
\label{ThermProp}
G(x) = \frac{1}{(2{\pi})^{d_s}} \, \Big(\frac{{\pi}}{\gamma}\Big)^{d_s/2} \sum^{\infty}_{n \, = \, - \infty} \frac{1}{x_n^{d_s/2}}
\, \exp \! \Big[ - \frac{{\vec x}^2}{4 \gamma x_n} - \mu H x_n \Big] \, \Theta (x_n) \, ,
\end{equation}
with
\begin{equation}
x_n \, \equiv \, x_4 + n \beta \, .
\end{equation}

We restrict ourselves to the infinite volume limit and evaluate the free energy density $z$, defined by
\begin{equation}
\label{freeEnergyDensity}
z = - \, T \, \lim_{L\to\infty} L^{-d_s} \, \ln \, [\mbox{Tr} \exp(-{\cal H}/T)] \, .
\end{equation}

From a conceptual point of view it is quite remarkable that the effective field theory method can be applied to one-dimensional systems,
such as ferromagnetic spin chains. In fact, ferromagnets represent a peculiar case. The crucial point is that, in the effective field
theory framework, the perturbative evaluation of the partition function is based on the suppression of loop diagrams by some power of
momentum. This suppression of loops depends on the spatial dimension $d_s$ of the system as well as on the dispersion relation of its
Goldstone bosons. Now for systems with a quadratic dispersion relation, such as the ferromagnet, each loop involves an integral of the
type
\begin{equation}
\int \! d \omega \, d^{d_s} k \ \frac{1}{\omega - \gamma {\vec k}^2} \ \propto \ p^{d_s} \, ,
\end{equation}
related to ferromagnetic magnons circling in the loop. On dimensional grounds the integral is proportional to $d_s$ powers of momentum.
While loops in three- (two-) dimensional ferromagnets are suppressed by three (two) powers of momentum, each loop in a Feynman diagram
referring to ferromagnetic spin chains is still suppressed by one power of momentum $p$. The one-loop diagram 3 is of order $p^3$, as it
involves ${\cal L}^2_{eff}$ ($p^2$) and one loop ($p$). The two-loop diagram 4 is of order $p^4$, as it involves one more loop compared to
diagram 3.

This suppression rule lies at the heart of the organization of the Feynman graphs of the partition function referring to ferromagnetic
spin chains depicted in Fig.~\ref{figure1}. Now we also understand why the piece ${\cal L}^6_{eff}$ is not needed for the present study:
the corresponding one-loop graph with a vertex from ${\cal L}^6_{eff}$ is of order $p^7$, i.e., beyond the order we are concerned with
here.

In the next section, we will evaluate the partition function of the one-dimensional ideal ferromagnet in full generality up to order
$p^4$. The evaluation of the partition function at order $p^5$ is much more involved. In particular, the renormalization and numerical
evaluation of the three-loop graph 5c turns out to be rather elaborate -- a detailed account of this calculation will be presented
elsewhere \citep{Hof12c}. Here we rather focus on the general structure of the low-temperature expansion and answer the question of which
contributions originate from noninteracting spin waves and which ones are due to the spin-wave interaction -- this question has never been
addressed so far. Still, we will also evaluate graph 5d which corresponds to noninteracting magnons, in order to compare our results with
the literature.

We emphasize that the suppression of loops in the case of ferromagnets is different from the loop suppression for systems with a linear
dispersion relation. There, a loop corresponds to an integral of the type
\begin{equation}
\int \! d \omega \, d^{d_s} k \ \frac{1}{{\omega}^2 - c^2 {\vec k}^2} \ \propto \ p^{d_s-1} \, .
\end{equation}
On dimensional grounds the integral is proportional to $d_s$-1 powers of momentum. This means that for antiferromagnetic magnons, loops in
three (two) spatial dimensions are suppressed by two (one) power of momentum and that the effective loop expansion perfectly works in
these cases. However, in one spatial dimension, loops are not suppressed at all and that's why the effective Lagrangian method cannot
be used to systematically analyze antiferromagnetic spin chains or any other one-dimensional (Pseudo-)Lorentz-invariant system in terms of
a loop expansion.

\section{Low-Temperature Properties of Ferromagnetic Spin Chains}
\label{Results}

We now consider those Feynman graphs depicted in Fig.~\ref{figure1} that contribute to the partition function up to order $p^4$ or,
equivalently, up to order $T^2$. We also include the diagram 5d, because we want to compare our results with the condensed matter
literature which is restricted to noninteracting spin waves. Additional information on finite-temperature effective Lagrangians and the
evaluation of the corresponding Feynman diagrams -- going beyond the outline given in the previous section -- can be found in
Ref.~\citep{Hof11a} (see section III and appendix A). Again we like to point out that we are considering one-dimensional {\it ideal}
ferromagnets, i.e., ferromagnetic spin chains which are governed by the isotropic exchange interaction between nearest neighbors and by
the interaction with a weak external magnetic field.

At leading order $p^2$, we have the tree graph 2 involving ${\cal L}^2_{eff}$, which merely leads to a temperature-independent
contribution to the free energy density,
\begin{equation}
z_{2} = - \Sigma \mu H \, .
\end{equation}

The leading temperature-dependent contribution is of order $p^3$ and stems from the one-loop graph 3. It is associated with a
$(d_s+1)$-dimensional nonrelativistic free Bose gas and amounts to
\begin{equation}
\label{z(3)T}
z^T_{3} = - \frac{1}{2 \pi^{\frac{1}{2}} \gamma^{\frac{1}{2}}} \, T^{\frac{3}{2}} \sum^{\infty}_{n=1} \frac{e^{- \mu H n \beta}}{n^{\frac{3}{2}}} \, .
\end{equation}

At order $p^4$, the first two-loop graph shows up. This contribution, related to graph 4, is proportional to single space derivatives of
the propagator at the origin,
\begin{equation}
\label{z(4)}
z_{4} \ \propto \, \Big[ {\partial}_{x_1} G(x) \Big]_{x=0} \, \Big[{\partial}_{x_1} G(x) \Big]_{x=0} = 0 \, ,
\end{equation}
and thus vanishes because the thermal propagator is invariant under parity, much like the Heisenberg Hamiltonian. Remember that the
effective Lagrangian - and therefore the thermal propagator -- inherits all the symmetries of the underlying Heisenberg model.

Finally, we include the one-loop graph 5d of order $p^5$, which corresponds to noninteracting magnons. Here, the next-to-leading order
Lagrangian ${\cal L}^4_{eff}$ comes into play through a two-magnon vertex,
\begin{equation}
\label{z(5d)}
z_{5d} = - \frac{2 \, l_3}{{\Sigma}} \, \Big[ {\partial}^4_{x_1} G(x) \Big]_{x=0} \, ,
\end{equation}
yielding the temperature-dependent contribution
\begin{equation}
\label{z(5d)T}
z^T_{5d} = - \frac{3 l_3}{4 {\pi}^{\frac{1}{2}} \Sigma {\gamma}^{\frac{5}{2}}} \, T^{\frac{5}{2}} \sum^{\infty}_{n=1}
\frac{e^{- \mu H n \beta}}{n^{\frac{5}{2}}} \, .
\end{equation}

Collecting terms, the free energy density of the ferromagnetic spin chain becomes
\begin{equation}
\label{FreeCollect}
z = - \Sigma \mu H - \frac{1}{2 \pi^{\frac{1}{2}} \gamma^{\frac{1}{2}}} \, T^{\frac{3}{2}} \, \sum^{\infty}_{n=1}
\frac{e^{- \mu H n \beta}}{n^{\frac{3}{2}}}
- \frac{3 l_3}{4 {\pi}^{\frac{1}{2}} \Sigma {\gamma}^{\frac{5}{2}}} \, T^{\frac{5}{2}} \sum^{\infty}_{n=1}
\frac{e^{- \mu H n \beta}}{n^{\frac{5}{2}}} + {\cal O}(p^5) \, .
\end{equation}
The contributions of order $T^{3/2}$ and $T^{5/2}$ arise from one-loop graphs and are both related to noninteracting spin waves. While the
former is exclusively determined by the leading-order effective constants $\Sigma$ and $F$ ($\gamma = F^2/\Sigma$), the latter involves
the next-to-leading-order coupling constant $l_3$.

It is quite remarkable that the spin-wave interaction does not yet manifest itself at next-to-leading order $p^4$ in the low-temperature
expansion of the free energy density. The only potential candidate, the two-loop diagram 4 of order $T^2$, turns out to be zero due to
parity. This is the analog of Dyson's statement that, in the case of the three-dimensional ideal ferromagnet, there is no term of order
$T^4$ in the low-temperature series of the free energy density. Likewise, there is no interaction term of order $T^3$ in the
low-temperature series of the free energy density referring to the two-dimensional ideal ferromagnet. Regardless of the spatial dimension,
the relevant two-loop diagram turns out to be zero due to parity \citep{Hof02,Hof11a,Hof12a,Hof12b}. In the case of ferromagnetic spin
chains, the spin-wave interaction enters through the three-loop graphs 5a, 5b, and 5c, yielding additional terms of order
$p^5 \propto T^{5/2}$ in the series (\ref{FreeCollect}).

It is important to stress that our rigorous approach is completely systematic and does not resort to any kind of approximations or ad hoc
assumptions. The structure of the above low-temperature series is an immediate consequence of the symmetries inherent in the
one-dimensional ideal ferromagnet.

In order to discuss the effect of a weak magnetic field, we expand the result (\ref{FreeCollect}) in the dimensionless parameter
\begin{equation}
\sigma = \mu H \beta = \frac{\mu H}{T} \, .
\end{equation}
Retaining all terms up to quadratic in $\sigma$, we obtain
\begin{eqnarray}
\label{FreeCollectSmallH}
z & = & - \Sigma \mu H - \frac{1}{2 \pi^{\frac{1}{2}} \gamma^{\frac{1}{2}}} \, T^{\frac{3}{2}} \, \Bigg\{ \zeta(\mbox{$\frac{3}{2}$})
- 2 \pi^{\frac{1}{2}} \sigma^{\frac{1}{2}} - \zeta(\mbox{$\frac{1}{2}$}) \sigma
+ \mbox{$\frac{1}{2}$}  \zeta(\mbox{$-\frac{1}{2}$}) \sigma^2
+ {\cal O}(\sigma^3) \Bigg\} \nonumber \\
& & - \frac{3 l_3}{4 {\pi}^{\frac{1}{2}} \Sigma {\gamma}^{\frac{5}{2}}} \, T^{\frac{5}{2}} \Bigg\{ \zeta(\mbox{$\frac{5}{2}$})
- \zeta(\mbox{$\frac{3}{2}$}) \sigma
+ \mbox{$\frac{4}{3}$} \pi^{\frac{1}{2}} \sigma^{\frac{3}{2}}
+ \mbox{$\frac{1}{2}$} \zeta(\mbox{$\frac{1}{2}$}) \sigma^2
+ {\cal O}(\sigma^3) \Bigg\}
+ {\cal O}(p^5) \, .
\end{eqnarray}
A thorough discussion of the range of validity of this series will be given in Sec.~\ref{MerminWagner}. As it turns out, it would be
inconsistent to take the limit $H \! \to \! 0$.

Let us also consider the low-temperature series for the energy density $u$, for the entropy density $s$, and for the heat capacity $c_V$
of the ferromagnetic spin chain. They are readily worked out from the thermodynamic relations
\begin{equation}
\label{Thermodynamics}
s = \frac{{\partial}P}{{\partial}T} \, , \qquad u = Ts - P \, , \qquad 
c_V = \frac{{\partial}u}{{\partial}T} = T \, \frac{{\partial}s}{{\partial}T} \, .
\end{equation}
Because the system is homogeneous, the pressure can be obtained from the temperature-dependent part of the free energy density,
\begin{equation}
\label{Pz}
P = z_0 - z \, ,
\end{equation}
such that the thermodynamic quantities amount to
\begin{eqnarray}
u & = & \frac{1}{2 \pi^{\frac{1}{2}} \gamma^{\frac{1}{2}}} \, T^{\frac{3}{2}} \, \Bigg\{ \sigma \sum^{\infty}_{n=1}
\frac{e^{- \sigma n}}{n^{\frac{1}{2}}}
+ \mbox{$\frac{1}{2}$} \sum^{\infty}_{n=1} \frac{e^{- \sigma n}}{n^{\frac{3}{2}}} \Bigg\} \nonumber \\
& & + \frac{3 l_3}{4 {\pi}^{\frac{1}{2}} \Sigma {\gamma}^{\frac{5}{2}}} \, T^{\frac{5}{2}} \, \Bigg\{ \sigma \sum^{\infty}_{n=1}
\frac{e^{- \sigma n}}{n^{\frac{3}{2}}} + \mbox{$\frac{3}{2}$} \sum^{\infty}_{n=1} \frac{e^{- \sigma n}}{n^{\frac{5}{2}}} \Bigg\}
+ {\cal O}(p^5) \, , \nonumber \\
s & = & \frac{1}{2 \pi^{\frac{1}{2}} \gamma^{\frac{1}{2}}} \, T^{\frac{1}{2}} \, \Bigg\{ \sigma \sum^{\infty}_{n=1}
\frac{e^{- \sigma n}}{n^{\frac{1}{2}}}
+ \mbox{$\frac{3}{2}$} \sum^{\infty}_{n=1} \frac{e^{- \sigma n}}{n^{\frac{3}{2}}} \Bigg\} \nonumber \\
& & + \frac{3 l_3}{4 {\pi}^{\frac{1}{2}} \Sigma {\gamma}^{\frac{5}{2}}} \, T^{\frac{3}{2}} \, \Bigg\{ \sigma \sum^{\infty}_{n=1}
\frac{e^{- \sigma n}}{n^{\frac{3}{2}}} + \mbox{$\frac{5}{2}$} \sum^{\infty}_{n=1} \frac{e^{- \sigma n}}{n^{\frac{5}{2}}} \Bigg\}
+ {\cal O}(p^3) \, , \\
c_V & = & \frac{1}{2 \pi^{\frac{1}{2}} \gamma^{\frac{1}{2}}} \, T^{\frac{1}{2}} \, \Bigg\{ {\sigma}^2 \sum^{\infty}_{n=1}
\frac{e^{- \sigma n}}{n^{-\frac{1}{2}}}
+ \sigma \sum^{\infty}_{n=1} \frac{e^{- \sigma n}}{n^{\frac{1}{2}}}
+ \mbox{$\frac{3}{4}$} \sum^{\infty}_{n=1} \frac{e^{- \sigma n}}{n^{\frac{3}{2}}} \Bigg\} \nonumber \\
& & + \frac{3 l_3}{4 {\pi}^{\frac{1}{2}} \Sigma {\gamma}^{\frac{5}{2}}} \, T^{\frac{3}{2}} \, \Bigg\{ {\sigma}^2 \sum^{\infty}_{n=1}
\frac{e^{- \sigma n}}{n^{\frac{1}{2}}} + 3 \sigma \sum^{\infty}_{n=1} \frac{e^{- \sigma n}}{n^{\frac{3}{2}}} + \mbox{$\frac{15}{4}$}
\sum^{\infty}_{n=1} \frac{e^{- \sigma n}}{n^{\frac{5}{2}}} \Bigg\}
+ {\cal O}(p^3) \, . \nonumber
\end{eqnarray}
Again, for a weak magnetic field $H$, the series may be expanded in the parameter $\sigma = \mu H /T$,
\begin{eqnarray}
u & = & \frac{1}{2 \pi^{\frac{1}{2}} \gamma^{\frac{1}{2}}} \, T^{\frac{3}{2}} \, \Big\{
\mbox{$\frac{1}{2}$} \zeta(\mbox{$\frac{3}{2}$})
+ \mbox{$\frac{1}{2}$} \zeta(\mbox{$\frac{1}{2}$}) \sigma
- \mbox{$\frac{3}{4}$} \zeta(\mbox{$-\frac{1}{2}$}) {\sigma}^2
+ {\cal O}({\sigma}^3) \Big\}  \nonumber \\
& & + \frac{3 l_3}{4 {\pi}^{\frac{1}{2}} \Sigma {\gamma}^{\frac{5}{2}}} \, T^{\frac{5}{2}} \, \Big\{
\mbox{$\frac{3}{2}$} \zeta(\mbox{$\frac{5}{2}$})
- \mbox{$\frac{1}{2}$} \zeta(\mbox{$\frac{3}{2}$}) \sigma
- \mbox{$\frac{1}{4}$} \zeta(\mbox{$\frac{1}{2}$}) {\sigma}^2
+ {\cal O}({\sigma}^3) \Big\}
+ {\cal O}(p^5) \, , \\
s & = & \frac{1}{2 \pi^{\frac{1}{2}} \gamma^{\frac{1}{2}}} \, T^{\frac{1}{2}} \, \Big\{
\mbox{$\frac{3}{2}$} \zeta(\mbox{$\frac{3}{2}$})
- 2 \sqrt{\pi} {\sigma}^{\frac{1}{2}}
- \mbox{$\frac{1}{2}$}\zeta(\mbox{$\frac{1}{2}$}) \sigma
- \mbox{$\frac{1}{4}$} \zeta(\mbox{$-\frac{1}{2}$}) {\sigma}^2
+ {\cal O}({\sigma}^3) \Big\}  \nonumber \\
& & + \frac{3 l_3}{4 {\pi}^{\frac{1}{2}} \Sigma {\gamma}^{\frac{5}{2}}} \, T^{\frac{3}{2}} \, \Big\{
\mbox{$\frac{5}{2}$} \zeta(\mbox{$\frac{5}{2}$})
- \mbox{$\frac{3}{2}$} \zeta(\mbox{$\frac{3}{2}$}) \sigma
+ \mbox{$\frac{4 \sqrt{\pi}}{3}$} {\sigma}^{\frac{3}{2}}
+ \mbox{$\frac{1}{4}$} \zeta(\mbox{$\frac{1}{2}$}) {\sigma}^2
+ {\cal O}({\sigma}^3) \Big\}
+ {\cal O}(p^3) \, ,  \nonumber
\end{eqnarray}
\begin{eqnarray}
c_V & = & \frac{1}{2 \pi^{\frac{1}{2}} \gamma^{\frac{1}{2}}} \, T^{\frac{1}{2}} \, \Big\{
\mbox{$\frac{3}{4}$} \zeta(\mbox{$\frac{3}{2}$})
+ \mbox{$\frac{1}{4}$} \zeta(\mbox{$\frac{1}{2}$}) \sigma
+ \mbox{$\frac{3}{8}$} \zeta(\mbox{$-\frac{1}{2}$}) {\sigma}^2
+ {\cal O}({\sigma}^3) \Big\} \nonumber \\
& & + \frac{3 l_3}{4 {\pi}^{\frac{1}{2}} \Sigma {\gamma}^{\frac{5}{2}}} \, T^{\frac{3}{2}} \, \Big\{
\mbox{$\frac{15}{4}$} \zeta(\mbox{$\frac{5}{2}$})
- \mbox{$\frac{3}{4}$} \zeta(\mbox{$\frac{3}{2}$}) \sigma
- \mbox{$\frac{1}{8}$} \zeta(\mbox{$\frac{1}{2}$}) {\sigma}^2
+ {\cal O}({\sigma}^3)\Big\} 
+ {\cal O}(p^3) \, ,
\end{eqnarray}
where we have retained terms up to quadratic in the magnetic field.

Note that all terms in the above series for $u$, $s$ and $c_V$ originate from the two one-loop graphs displayed in Fig.~\ref{figure1}. The
explicit contribution due to the spin-wave interaction, entering at order $p^5 \propto T^{5/2}$ ($p^3 \propto T^{3/2}$) for $u$ ($s, c_V$),
will be considered in detail in Ref.~\citep{Hof12c}. Here we want to emphasize that there is no interaction term of order
$p^4 \propto T^2$ in the energy density and no interaction term of order $p^2 \propto T$ in the entropy density and heat capacity.

Let us now turn to the magnetization. With the expression for the free energy density (\ref{FreeCollect}), the low-temperature expansion
for the magnetization 
\begin{equation}
\Sigma(T,H) \, = \, - \frac{\partial z}{\partial(\mu H)}
\end{equation}
of ferromagnetic spin chains takes the form
\begin{equation}
\label{magnetizationD1}
\frac{\Sigma(T,H)}{\Sigma} \; = \; 1 - {\tilde \alpha}_0 \, T^{\frac{1}{2}} - {\tilde \alpha}_1 \, T^{\frac{3}{2}} + {\cal O}(p^3) \, .
\end{equation}
The coefficients $\tilde \alpha_i$ depend on the dimensionless ratio $\sigma = \mu H/T$ and are given by
\begin{eqnarray}
\label{SigmaCollectT(H=0)}
{\tilde \alpha}_0 & = & \frac{1}{2 {\pi}^{\frac{1}{2}} \Sigma {\gamma}^{\frac{1}{2}}} \, \sum^{\infty}_{n=1} \frac{e^{- \sigma n}}{n^{\frac{1}{2}}}
\, , \nonumber \\
{\tilde \alpha}_1 & = & \frac{3 l_3}{4 {\pi}^{\frac{1}{2}} {\Sigma}^2 {\gamma}^{\frac{5}{2}}} \, \sum^{\infty}_{n=1}
\frac{e^{- \sigma n}}{n^{\frac{3}{2}}} \, .
\end{eqnarray}

Finally, the susceptibility,
\begin{equation}
\chi(T,H) \, = \, \frac{\partial \Sigma(T,H)}{\partial(\mu H)} \, ,
\end{equation}
of ferromagnetic spin chains amounts to
\begin{equation}
\chi(T,H) \; = \; {\tilde \kappa}_0 \, T^{-\frac{1}{2}} + {\tilde \kappa}_1 \, T^{\frac{1}{2}} + {\cal O}(p) \, ,
\end{equation}
with coefficients
\begin{eqnarray}
{\tilde \kappa}_0 & = & \frac{1}{2 {\pi}^{\frac{1}{2}} {\gamma}^{\frac{1}{2}}} \, \sum^{\infty}_{n=1} \frac{e^{- \sigma n}}{n^{-\frac{1}{2}}}
\, , \nonumber \\
{\tilde \kappa}_1 & = & \frac{3 l_3}{4 {\pi}^{\frac{1}{2}} {\Sigma} {\gamma}^{\frac{5}{2}}} \, \sum^{\infty}_{n=1}
\frac{e^{- \sigma n}}{n^{\frac{1}{2}}} \, .
\end{eqnarray}
In what follows, we will critically examine the range of validity of the series presented in this section and thereby put the above
low-temperature expansions for a one-dimensional system on a firm basis -- on the same footing as the low-temperature series for ferro-
and antiferromagnets in three or two spatial dimensions.

\section{Low-Temperature Series: Range of Validity}
\label{MerminWagner}

The range of validity of the low-temperature series presented in this work is restricted due to the Mermin-Wagner theorem \citep{MW68}.
The theorem states that, in one or two spatial dimensions, no spontaneous symmetry breaking at any finite temperature can occur in the
O(3)-invariant Heisenberg model. In the context of the ferromagnet an energy gap is generated nonperturbatively and the correlation
length of the magnons no longer is infinite. Still, even in one spatial dimension, the correlation length is large, being proportional to
the inverse temperature \citep{Kop89},
\begin{equation}
\label{npcorrelation}
\xi_{np} =  a \, C^{(0)}_{\xi} \, \frac{J S^2}{T} \, \Big[1 + C^{(1)}_{\xi} \frac{1}{\pi} \sqrt{\frac{T}{J S^3}} + {\cal O}(T)  \Big] \, .
\end{equation}
Here $a$ is the spacing between two neighboring sites of the spin chain and the quantities $C^{(0)}_{\xi}$ and $C^{(1)}_{\xi}$ are
dimensionless constants. According to Ref.~\citep{Kop89}, they take the values
\begin{equation}
C^{(0)}_{\xi} =  1.14 \pm 0.11 \, , \qquad C^{(1)}_{\xi} = 0.6514 \pm 0.0012 \, .
\end{equation}
On the other hand, the value for the first constant, quoted in Refs.~\citep{AA88,Tak87a}, is $C^{(0)}_{\xi} = 1$, which is the number we
will use for the estimate below. It is important to note that, unlike in two spatial dimensions \citep{KC89}, the correlation length of
magnons in one spatial dimension no longer is exponentially large. 

Apart from the nonperturbatively generated correlation length $\xi_{np}$, ferromagnetic magnons are also characterized by the correlation
length $\xi$ which is related to the magnetic field. As long as the correlation length $\xi$ of the Goldstone bosons is much smaller than
the nonperturbatively generated correlation length $\xi_{np}$, our low-temperature series are valid: in this regime, the spin-waves are
well-defined and represent the relevant low-energy degrees of freedom. A natural way to define the correlation length $\xi$ for
ferromagnetic magnons is based on their dispersion relation,
\begin{equation}
\omega(k) = \gamma k^2 + \mu H + {\cal O}(k^4) \, , \quad \gamma \equiv \frac{F^2}{\Sigma} \, ,
\end{equation}
leading to
\begin{equation}
\label{correlation}
\xi = \sqrt{\frac{\gamma}{\mu H}} = \frac{F}{\sqrt{\Sigma\mu H}} \, .
\end{equation}
This quantity has dimension of length and tends to infinity if the magnetic field is switched off. Indeed, the correlation length of the
magnons in a ferromagnetic spin chain is infinite at zero temperature.

For the low-temperature series to be valid, the ratio
\begin{equation}
x = \xi / \xi_{np}
\end{equation}
of the two correlation lengths must be a small number. Using
Eqs.~(\ref{npcorrelation}) and (\ref{correlation}), and expressing the effective constant $\gamma$ in terms of the exchange integral $J$
of the underlying theory as \citep{Kit96}
\begin{equation}
\label{gammaMatch}
\gamma = J S a^2 \, ,
\end{equation}
we arrive at
\begin{equation}
\label{npestimate}
\frac{\mu H}{T} = \frac{1}{S^3 x^2} \, \frac{T}{J} \, .
\end{equation}
For $S=\mbox{$\frac{1}{2}$}$ and with the ratio $x=\mbox{$\frac{1}{10}$}$, we obtain the relation
\begin{equation}
\label{ratios}
\frac{\mu H}{T} = 800 \, \frac{T}{J} \, ,
\end{equation}
which can be interpreted as follows. The exchange integral $J$ defines a scale in the underlying theory and for the effective expansion to
be valid, the temperature has to be small with respect to this scale. For practical purposes we may choose
\begin{equation}
\frac{T}{J} = \frac{1}{50} \quad \Longrightarrow \quad \frac{\mu H}{T} = \sigma = 16 \, .
\end{equation}
The parameter $\sigma$ can thus be large, i.e., the magnetic field need not be small compared to the temperature. What is essential,
however, is that the magnetic field itself -- much like the temperature -- is small compared to the intrinsic scale $J$ of the underlying
theory.

It is important to note that we cannot completely switch off the magnetic field in our low-temperature expansions. Rather, we start
running into trouble as soon as we choose a ratio $\mu H /T$, which is smaller than $800 \; T/J$: we then leave the domain of validity of
the low-temperature series derived in this work, because the effective calculation does not take into account the nonperturbative effect.

\begin{figure}
\includegraphics[width=12cm]{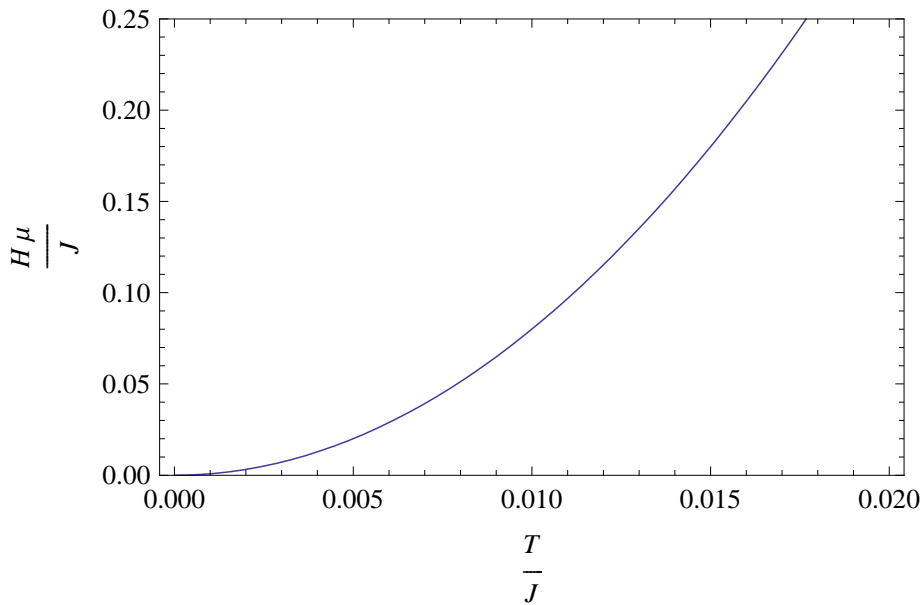}

\caption{Range of validity of the effective low-temperature expansions of the ferromagnetic spin chain. The allowed parameter regime
 corresponds to the area above the curve and is restricted to very low temperatures.}
\label{figure2}
\end{figure} 

To illustrate the range of validity, we consider the two-dimensional domain defined by the parameters $T/J$ and $\mu H/J$, which both have
to be small for the effective expansion to be valid. In terms of these parameters the condition (\ref{ratios}) takes the form
\begin{equation}
\frac{\mu H}{J} = 800 \, \frac{T^2}{J^2} \, .
\end{equation}
This is the line plotted in Fig.~\ref{figure2}. In the parameter space above that curve, the low-temperature series derived in this work
are valid. In one spatial dimension, the parameter regime is thus quite restricted. In particular, note that the horizontal axis
corresponding to zero magnetic field is outside the allowed domain.

It is very instructive to compare this result with the range of validity of the analogous low-temperature series, referring to
two-dimensional ideal ferromagnets. There, the nonperturbatively generated correlation length is exponentially large, the argument of the
exponential being proportional to the inverse temperature \citep{KC89},
\begin{equation}
\label{npcorrelationD2}
\xi_{np} = C_{\xi} a S^{-\frac{1}{2}} \, \sqrt{\frac{T}{J S^2}} \, \exp \! \Big[\frac{2 \pi J S^2}{T} \Big]  \qquad (d_s = 2) \, ,
\end{equation}
where $a$ is the spacing between two neighboring sites on the square lattice, and the quantity $ C_{\xi} \approx 0.05$ is a dimensionless
constant.

Following the same steps as before, for two-dimensional ideal ferromagnets one derives the relation \citep{Hof12a}
\begin{equation}
\label{npestimateD2}
\frac{\mu H}{T} = \frac{400 S^4}{x^2} \, \frac{J^2}{T^2} \, \exp\Big[-\frac{4 \pi J S^2}{T}\Big] \qquad (d_s=2) \, ,
\end{equation}
or
\begin{equation}
\label{ratiosD2}
\frac{\mu H}{T} = 2500 \, \frac{J^2}{T^2} \, \exp\Big[- \pi \frac{J}{T}\Big] \qquad (d_s=2) \, ,
\end{equation}
for $S=\mbox{$\frac{1}{2}$}$ and $x=\mbox{$\frac{1}{10}$}$. In two spatial dimensions it is also conceptually inconsistent to switch off
the magnetic field: here we start running into trouble as soon as the ratio $\mu H /T$ is smaller than the value given by the RHS of
Eq.~(\ref{ratiosD2}).

\begin{figure}
\includegraphics[width=12cm]{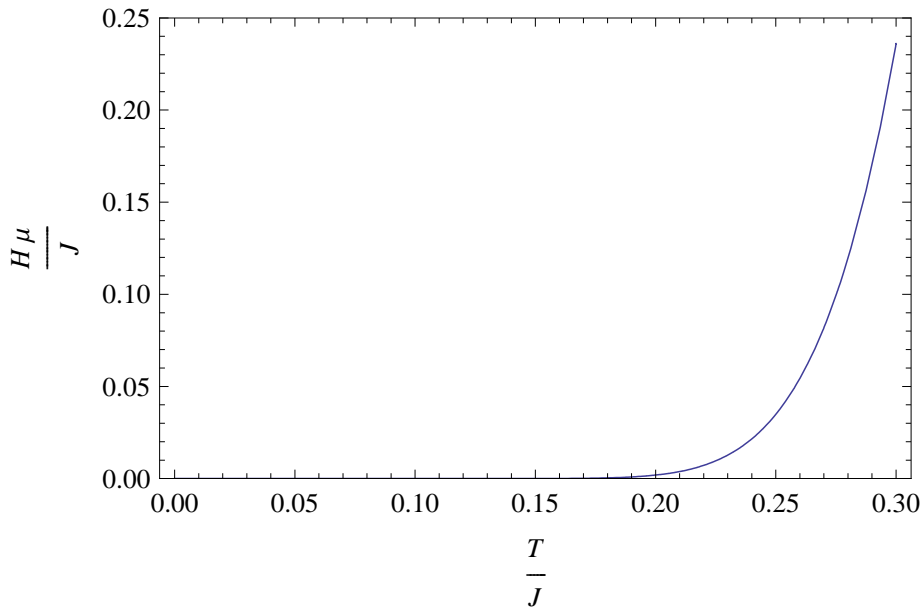}

\caption{Range of validity of the effective low-temperature expansions of the two-dimensional ideal ferromagnet. The allowed parameter
regime corresponds to the area above the curve.}
\label{figure3}
\end{figure} 

Rewritten in terms of the parameters $T/J$ and $\mu H/J$, the condition (\ref{ratiosD2}) implies
\begin{equation}
\frac{\mu H}{J} = 2500 \, \frac{J}{T} \, \exp\Big[- \pi \frac{J}{T}\Big] \qquad (d_s=2) \, .
\end{equation}
This is the line plotted in Fig.~\ref{figure3}, indicating that the effective low-temperature series derived in
Refs.~\citep{Hof12a,Hof12b} are valid in the parameter space above this line. As one can see, the allowed parameter regime is much larger
than in one spatial dimension. Again, this is due to the fact that in two spatial dimensions, the nonperturbative correlation length is
exponentially large.

\begin{figure}
\includegraphics[width=12cm]{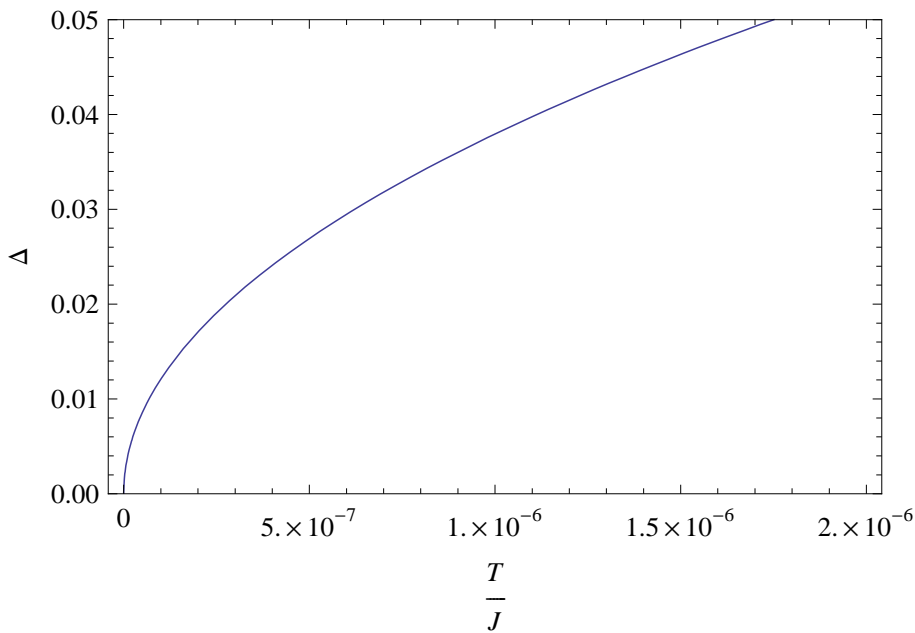}

\caption{Ferromagnetic spin chain: The Taylor expansion for the free energy density in the parameter $\sigma$,
Eq.~(\ref{FreeCollectSmallH}), only makes sense at very low temperatures compared to the scale $J$.}
\label{figure4}
\end{figure}

In the previous section we have considered the effect of a weak magnetic field by Taylor expanding the low-temperature series for
$z, u, s$ and $c_V$ in the small parameter $\sigma = \mu H/T$. In order to illustrate in which parameter regime the Taylor expansion does
make sense for the free energy density, in Fig.~\ref{figure4}, we have plotted the quantity $\Delta$
\begin{equation}
\label{DeltaDim1}
\Delta = 1 - \frac{\sum^{\infty}_{n=1} \frac{e^{- \mu H n \beta}}{n^{3/2}}}{\zeta(\mbox{$\frac{3}{2}$})} \, , \qquad 
\sigma = \mu H \beta = 800 \, \frac{T}{J} \, ,
\end{equation}
as a function of the parameter $T/J$. The quantity $\Delta$ must be small, let us say $\Delta \leq 0.05$, in order for the Taylor series
(\ref{FreeCollectSmallH}) to make sense: the leading term proportional to $\zeta(\mbox{$\frac{3}{2}$})$ in the expansion
(\ref{FreeCollectSmallH}) then makes up 95 \% or more of the full leading order contribution displayed in the numerator of
Eq.~(\ref{DeltaDim1}). However, according to Fig.~\ref{figure4}, for that to be the case, the temperature has to be extremely small
compared to the scale $J$. We thus conclude that the Taylor expansions for $z$, as well for $u, s$ and $c_V$, in the parameter $\sigma$
only make sense in a very small domain. Still, we have provided these expansions in the previous section for completeness, and also
because in the next section, we want to compare them with the literature. We have to stress, however, that their range of validity has
never been thoroughly discussed in the literature, i.e., the leading terms in these Taylor series were given in
Refs.~\citep{Tak86,Tak87a,KSK03}, without actually pointing out that these series are only valid in an extremely small parameter regime.

\begin{figure}
\includegraphics[width=12cm]{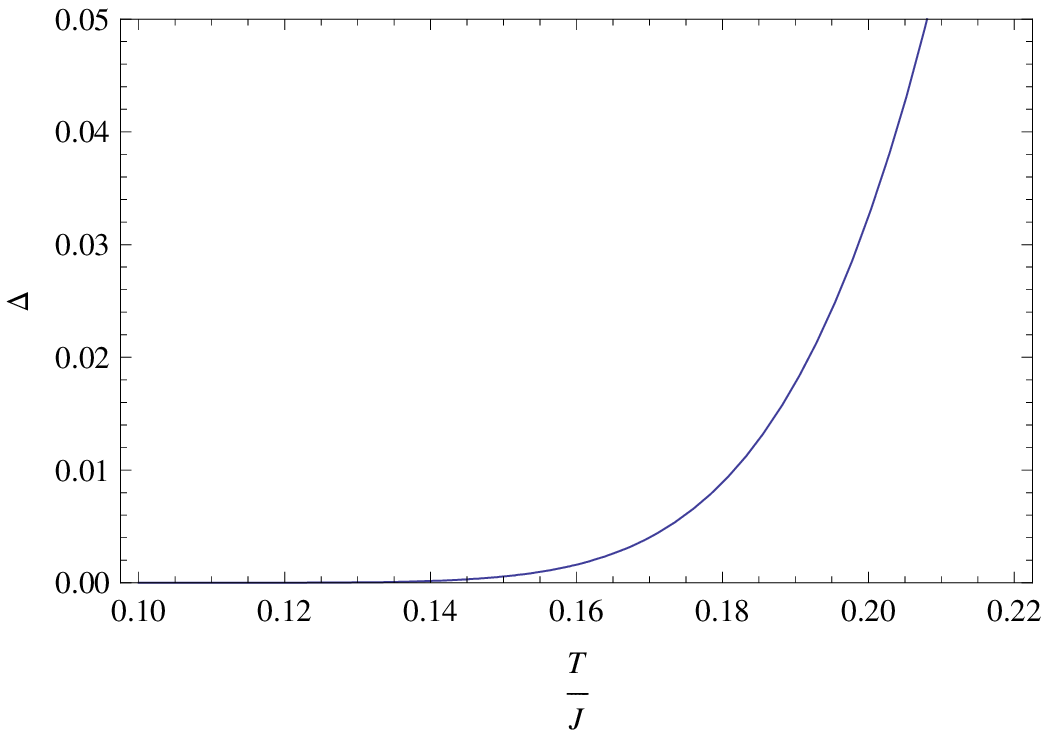}

\caption{Two-dimensional ideal ferromagnet: The Taylor expansion for the free energy density in the parameter $\sigma$,
Eq.~(III.8) of Ref.~\citep{Hof12a}, makes sense up to temperatures which are about one fifth of the scale $J$.}
\label{figure5}
\end{figure} 

Again it is instructive to also consider the situation in two-dimensional ideal ferromagnets. The analogous quantity $\Delta$ for the
free energy density, according to Eq.~(III.8) of Ref.~\citep{Hof12a}, is
\begin{equation}
\label{DeltaDim2}
\Delta = 1 - \frac{\sum^{\infty}_{n=1} \frac{e^{- \mu H n \beta}}{n^2}}{\zeta(2)} \, ,
\qquad \sigma = \mu H \beta = 2500 \, \frac{J^2}{T^2} \, \exp\Big[- \pi \frac{J}{T}\Big] \qquad \ (d_s=2) \, .
\end{equation}
This is the curve plotted in Fig.~\ref{figure5}, indicating that the situation here is entirely different: in two spatial dimensions,
where the nonperturbative correlation length is exponentially large, the parameter regime in which the Taylor expansion in $\sigma$ for
$z$, as well as for $u, s$ and $c_V$ makes sense, is much larger: the temperature need not be tiny with respect to the scale $J$. 

We have argued that the magnetic field cannot be switched off in the present study. While the case ${\vec H} = 0$ is beyond the reach of
the effective expansion presented here, it is not beyond the reach of the effective field theory method. Rather, one has to establish a
different type of systematic perturbative expansion, which can cope with a zero magnetic field. Work in this direction is in progress.

\section{Comparison with the Literature}
\label{CDM}

The thermodynamic properties of ferromagnetic spin chains have been analyzed with a variety of microscopic methods: spin-wave theory,
Bethe ansatz, Schwinger-boson mean-field theory, Green functions and renormalization group methods. Almost all of these references obtain
a term of order $T^2$ in the low-temperature expansion of the free energy density, which appears to be in contradiction with the
systematic effective field theory calculation. In fact, we have argued that there is no $T^2$-term in this series, because the only
potential candidate, the two-loop graph 4 of Fig.~\ref{figure1}, vanishes due to parity. This is an exact statement and represents the
analog of Dyson's statement that, in three spatial dimensions, there is no term of order $T^4$ in the free energy density of an ideal
ferromagnet.

The essential point is to realize that most of the above references refer to ferromagnetic spin chains in zero external magnetic field.
This is not the domain where the effective Lagrangian method presented in this work operates: for our expansions to be valid, the magnetic
field is always different from zero.

While the focus of the pioneering articles by Takahashi, Refs.~\citep{Tak86,Tak87a}, is on $H = 0$, the nonzero field case is also
considered there. The method developed and advocated is modified spin-wave theory. As Takahashi states in Ref.~\citep{Tak86}, modified
spin-wave theory is restricted to zero external magnetic field. Conventional spin-wave theory, on the other hand, should be used if one
wants to study the low-temperature properties of ferromagnetic spin chains in the presence of a magnetic field. Indeed, conventional
spin-wave theory predicts that there is no term of order $T^2$ in the free energy density of a one-dimensional ferromagnet
\citep{Tak86,Tak87a},
\begin{equation}
\label{TakahashiConventionalSWT}
z = - T \, \Bigg\{ 1.0421869 { \Big( \frac{T}{J} \Big)}^{\frac{1}{2}} + 0.0668971 {\Big( \frac{T}{J} \Big)}^{\frac{3}{2}} + \dots  \Bigg\}
\qquad (S = \mbox{$\frac{1}{2}$}) \, ,
\end{equation}
in agreement with the systematic effective field theory result. However, two comments are in order here.

First, Takahashi's analysis is restricted to noninteracting spin waves -- in accordance with the effective analysis, free magnon particles
do not produce a term of order $T^2$ in the above series. The crucial point is that, as the effective analysis demonstrates, the spin-wave
interaction does not lead to a $T^2$- term, either. This result is entirely new.

Second, Eq.~(\ref{TakahashiConventionalSWT}) -- in view of the way it was derived -- cannot be trusted. Apparently, in order to obtain the
above expression, Takahashi has taken the limit $H \to 0$, which appears to be conceptually inconsistent: conventional spin-wave theory
does not operate in this sector. Likewise, as we have discussed at length in the previous section, the magnetic field in our effective
low-temperature series cannot be switched off, either. Furthermore, the statements that -- in the absence of a magnetic field --
conventional spin-wave theory {\it applies to some degree} (see Ref.~\citep{Tak87a}, p.~168), or that conventional spin-wave theory {\it
is valid in some sense} (see Ref.~\citep{Tak05}, p.~156), are misleading in our opinion.

Takahashi's expression (\ref{TakahashiConventionalSWT}) can still be used to extract the effective low-energy coupling $l_3$ by taking the
same (conceptually inconsistent) limit $H \to 0$ in our effective expansion (\ref{FreeCollectSmallH}). Matching the two expressions, we
end up with
\begin{equation}
\label{l3MatchS12}
l_3 =  \frac{\sqrt{\pi} c_{5/2}}{6 \sqrt{2} \zeta(\mbox{$\frac{5}{2}$})} \, J a^3 \approx  0.0104 \, J a^3 \qquad
(S = \mbox{$\frac{1}{2}$}) \, ,
\end{equation}
where the quantity $c_{5/2} = 0.0668971$ is the second coefficient in Takahashi's expansion (\ref{TakahashiConventionalSWT}). For general
spin $S$, the effective constant $l_3$ reads
\begin{equation}
\label{l3Match}
l_3 =  \frac{J S^2 a^3}{24} \, .
\end{equation}

The ferromagnetic spin chain in nonzero magnetic field was also considered in Ref.~\citep{KSK03}, where another variant of conventional
spin-wave theory, capable to deal with low-dimensional systems -- spin-wave theory at constant order parameter -- was invented. The
authors also discuss the case $H \neq 0$ and obtain the following expansion for the magnetization $m(H)$: 
\begin{equation}
\label{KopietzSWT}
\frac{m(H)}{S} = 1- \frac{\zeta(\mbox{$ \frac{1}{2} $})}{2 S \sqrt{\pi}} \, \sqrt{t} - \frac{1}{2 S} \, \sqrt{\frac{t}{v}}
+ {\cal O}(t, t^{3/2}v^{-1/2}) \, , \qquad t = \frac{T}{J S} \, , \ v = \frac{H}{T} \, .
\end{equation}
Indeed, Eq.~(\ref{KopietzSWT}) agrees with the leading terms of our effective result (\ref{magnetizationD1}).

For the reader's convenience, expressing the effective constants $\gamma$ and $l_3$ in terms of microscopic parameters according to
Eq.~(\ref{gammaMatch}) and Eq.~(\ref{l3Match}), we rewrite our series in a form where the $1/S$ expansion becomes manifest: 
\begin{eqnarray}
z & = & -  \frac{S \mu H}{a} - \frac{1}{2 \pi^{\frac{1}{2}} \sqrt{J S} a} \, T^{\frac{3}{2}} \, \sum^{\infty}_{n=1}
\frac{e^{- \mu H n \beta}}{n^{\frac{3}{2}}} \nonumber \\
& & - \frac{1}{32 {\pi}^{\frac{1}{2}} \sqrt{J^3 S^3} a} \, T^{\frac{5}{2}} \sum^{\infty}_{n=1}
\frac{e^{- \mu H n \beta}}{n^{\frac{5}{2}}} + {\cal O}(p^5) \, , \nonumber \\
u & = & \frac{1}{2 \pi^{\frac{1}{2}} \sqrt{J S} a} \, T^{\frac{3}{2}} \, \Bigg\{ \sigma \sum^{\infty}_{n=1}
\frac{e^{- \sigma n}}{n^{\frac{1}{2}}} + \mbox{$\frac{1}{2}$} \sum^{\infty}_{n=1} \frac{e^{- \sigma n}}{n^{\frac{3}{2}}} \Bigg\} \nonumber \\
& & +  \frac{1}{32 {\pi}^{\frac{1}{2}} \sqrt{J^3 S^3} a} \, T^{\frac{5}{2}} \, \Bigg\{ \sigma \sum^{\infty}_{n=1}
\frac{e^{- \sigma n}}{n^{\frac{3}{2}}} + \mbox{$\frac{3}{2}$} \sum^{\infty}_{n=1} \frac{e^{- \sigma n}}{n^{\frac{5}{2}}} \Bigg\}
+ {\cal O}(p^5) \, , \nonumber \\
s & = & \frac{1}{2 \pi^{\frac{1}{2}} \sqrt{J S} a} \, T^{\frac{1}{2}} \, \Bigg\{ \sigma \sum^{\infty}_{n=1}
\frac{e^{- \sigma n}}{n^{\frac{1}{2}}}
+ \mbox{$\frac{3}{2}$} \sum^{\infty}_{n=1} \frac{e^{- \sigma n}}{n^{\frac{3}{2}}} \Bigg\} \nonumber \\
& & +  \frac{1}{32 {\pi}^{\frac{1}{2}} \sqrt{ J^3 S^3} a} \, T^{\frac{3}{2}} \, \Bigg\{ \sigma \sum^{\infty}_{n=1}
\frac{e^{- \sigma n}}{n^{\frac{3}{2}}} + \mbox{$\frac{5}{2}$} \sum^{\infty}_{n=1} \frac{e^{- \sigma n}}{n^{\frac{5}{2}}} \Bigg\}
+ {\cal O}(p^3) \, , \nonumber \\
c_V & = & \frac{1}{2 \pi^{\frac{1}{2}} \sqrt{J S} a} \, T^{\frac{1}{2}} \, \Bigg\{ {\sigma}^2 \sum^{\infty}_{n=1}
\frac{e^{- \sigma n}}{n^{-\frac{1}{2}}}
+ \sigma \sum^{\infty}_{n=1} \frac{e^{- \sigma n}}{n^{\frac{1}{2}}}
+ \mbox{$\frac{3}{4}$} \sum^{\infty}_{n=1} \frac{e^{- \sigma n}}{n^{\frac{3}{2}}} \Bigg\} \nonumber \\
& & +  \frac{1}{32 {\pi}^{\frac{1}{2}} \sqrt{J^3 S^3} a} \, T^{\frac{3}{2}} \, \Bigg\{ {\sigma}^2 \sum^{\infty}_{n=1}
\frac{e^{- \sigma n}}{n^{\frac{1}{2}}} + 3 \sigma \sum^{\infty}_{n=1} \frac{e^{- \sigma n}}{n^{\frac{3}{2}}} + \mbox{$\frac{15}{4}$}
\sum^{\infty}_{n=1} \frac{e^{- \sigma n}}{n^{\frac{5}{2}}} \Bigg\}
+ {\cal O}(p^3) \, , \nonumber
\end{eqnarray}
\begin{eqnarray}
\frac{\Sigma(T,H)}{\Sigma} & = & 1 - \frac{1}{2 \pi^{\frac{1}{2}} \sqrt{J S^3}}\, \sum^{\infty}_{n=1}
\frac{e^{- \sigma n}}{n^{\frac{1}{2}}}\, T^{\frac{1}{2}}
-  \frac{1}{32 {\pi}^{\frac{1}{2}} \sqrt{J^3 S^5}} \, \sum^{\infty}_{n=1}
\frac{e^{- \sigma n}}{n^{\frac{3}{2}}} \, T^{\frac{3}{2}} + {\cal O}(p^3) \, , \nonumber \\
\chi(T,H) & = &  \frac{1}{2 \pi^{\frac{1}{2}} \sqrt{J S}  a} \, \sum^{\infty}_{n=1} \frac{e^{- \sigma n}}{n^{-\frac{1}{2}}}
\, T^{-\frac{1}{2}}
+ \frac{1}{32 {\pi}^{\frac{1}{2}} \sqrt{J^3 S^3} a} \, \sum^{\infty}_{n=1}
\frac{e^{- \sigma n}}{n^{\frac{1}{2}}} \, T^{\frac{1}{2}} + {\cal O}(p) \, .
\end{eqnarray}
In contrast to Refs.~\citep{Tak86,Tak87a,KSK03}, we have carefully discussed the range of validity of the above series in the previous
section, and thus have put these low-temperature expansions on safe grounds -- on the same footing as the low-temperature series for
ferro- and antiferromagnets in three and two spatial dimensions.

It is important to emphasize that all the theoretical references providing low-temperature series for ferromagnetic spin chains in the
presence of a magnetic field, were restricted to {\it free} magnons so far. The important question of whether the spin-wave interaction
-- in the case ${\vec H} \neq 0$ -- already shows up at order $T^{5/2}$ in the free energy density, or rather beyond, has never been
discussed. In other words, it remained unclear whether the series in Refs.~\citep{Tak86,Tak87a,KSK03}, referring to the ideal magnon gas
and derived within spin-wave theory, are indeed complete up to order $T^{5/2}$.

We have demonstrated that the spin-wave interaction starts manifesting itself at the three-loop level in the systematic effective
expansion. In the low-temperature series of the free energy density, the corresponding three-loop graphs 5a, 5b, and 5c of
Fig.~\ref{figure1}, indeed lead to a contribution of order $p^5 \propto T^{5/2}$. The explicit evaluation is quite involved and will be
presented elsewhere \citep{Hof12c}. Here we rather wanted to draw the attention to the general structure of the low-temperature series in
the presence of a weak external magnetic field and critically examine their range of validity, as well as compare our systematic results
with the condensed matter literature which is restricted to noninteracting spin waves.

\section{Conclusions}
\label{Summary}

We have studied the low-temperature behavior of ferromagnetic spin chains in the presence of a weak external magnetic field. While these
systems have been investigated by many authors using different techniques, such as Bethe ansatz, spin-wave theory and Schwinger-boson
mean-field theory, in the present study we have made use of the systematic effective Lagrangian method. We have evaluated the
low-temperature expansion of the partition function of ferromagnetic spin chains in a weak magnetic field up to two-loop order, and
derived the low-temperature series for the energy density, entropy density, heat capacity, magnetization and susceptibility.
Interestingly, the spin-wave interaction does not yet manifest itself at this order in the low-temperature expansions -- the only two-loop
graph turns out to be zero due to parity. The spin-wave interaction only enters at the three-loop level.

From a conceptual point of view, it is the first time that the low-temperature properties of systems defined in one spatial dimension have
been analyzed within the systematic effective loop expansion. One-dimensional systems which display a linear, i.e., relativistic,
dispersion law cannot be systematically analyzed with effective Lagrangians, because the loop counting breaks down. However, for systems
with a quadratic dispersion relation like the ferromagnet, the method perfectly works, because loop graphs are still suppressed by one
power of momentum.

We have carefully examined the range of validity of the low-temperature series which is quite restricted. Unlike in two spatial
dimensions, where the nonperturbatively generated correlation length of the spin waves is exponentially large at low temperatures, in one
spatial dimension, the nonperturbative correlation length is only proportional to the inverse temperature. As a consequence, both in one
and two spatial dimensions, it is conceptually inconsistent to switch off the magnetic field in our series as we would then leave their
domain of validity. We have confronted our results with those obtained by spin-wave theory and have pointed out that there are some
misleading statements in the literature.

In the above microscopic studies, the magnons were considered as ideal Bose particles -- the problem of the spin-wave interaction was
neglected and it thus remained unclear whether the low-temperature series given in these articles are complete or will receive additional
corrections due to the interaction.  While we have argued that, in the presence of a magnetic field, the spin-wave interaction enters at
order $p^5 \propto T^{5/2}$ in the free energy density of ferromagnetic spin chains, the explicit evaluation of the corresponding
three-loop graphs has not been considered here -- this will be the subject of Ref.~\citep{Hof12c}.

The present study shows that the effective Lagrangian method is a very powerful tool to analyze the general structure of the
low-temperature expansion of the partition function for systems exhibiting collective magnetic behavior. Not only have we rigorously
discussed the effect of the spin-wave interaction and a weak external magnetic field in a systematic manner, but also have we put our
low-temperature series on a firm basis.

It would be very interesting to establish the effective Lagrangian method in the parameter regime where the magnetic field is zero. This
quite nontrivial problem has been solved in Ref.~\citep{HN93} for the two-dimensional antiferromagnet in zero magnetic and staggered
field. Work on transferring these techniques to the ferromagnet is in progress. 

\section*{Acknowledgments}
The author would like to thank D.\ Dmitriev for correspondence and U.-J. Wiese for useful comments regarding the manuscript.

\end{document}